\documentclass[aps,prl,groupedaddress,reprint,superscriptaddress]{revtex4-2}
\usepackage{amsfonts}
\usepackage{amsmath}\usepackage{amssymb}
\usepackage{graphicx}% Include figure files
\usepackage{dcolumn}% Align table columns on decimal point
\usepackage{bm}% bold math
\usepackage{color}
\usepackage[normalem]{ulem}
\usepackage{verbatim}
\usepackage{dutchcal}
\usepackage{ulem}
\newcommand{\cz}[1]{\textcolor{black}{#1}}

\begin{document}

\title{Why does dissolving salt in water decrease its dielectric permittivity}

%\date{\today}% It is always \today, today,
             %  but any date may be explicitly specified
             
\author{Chunyi Zhang}
\affiliation{Department of Physics, Temple University, Philadelphia, Pennsylvania 19122, USA}

\author{Shuwen Yue}
\affiliation{Department of Chemical and Biological Engineering, Princeton University, Princeton, New Jersey 08544, USA}

\author{Athanassios Z. Panagiotopoulos}
\affiliation{Department of Chemical and Biological Engineering, Princeton University, Princeton, New Jersey 08544, USA}

\author{Michael L. Klein}
\email{mlklein@temple.edu}
\affiliation{Department of Physics, Temple University, Philadelphia, Pennsylvania 19122, USA}
\affiliation{Institute for Computational Molecular Science, Temple University, Philadelphia, Pennsylvania 19122, USA}
\affiliation{Department of Chemistry, Temple University, Philadelphia, Pennsylvania 19122, USA}

\author{Xifan Wu}
\email{xifanwu@temple.edu}
\affiliation{Department of Physics, Temple University, Philadelphia, Pennsylvania 19122, USA}
\affiliation{Institute for Computational Molecular Science, Temple University, Philadelphia, Pennsylvania 19122, USA}

%\date{\today}% It is always \today, today,
             %  but any date may be explicitly specified

\begin{abstract}
The dielectric permittivity of salt water decreases on dissolving more salt. For nearly a century, this phenomenon has been explained by invoking saturation in the dielectric response of the solvent water molecules. Herein, we employ an advanced deep neural network (DNN), built using data from density functional theory, to study the dielectric permittivity of sodium chloride solutions. Notably, the decrease in the dielectric permittivity as a function of concentration, computed using the DNN approach, agrees well with experiments. Detailed analysis of the computations reveals that the dominant effect, caused by the intrusion of ionic hydration shells into the solvent hydrogen-bond network, is the disruption of dipolar correlations among water molecules. Accordingly, the observed decrease in the dielectric permittivity is mostly due to increasing suppression of the collective response of solvent waters. 
\end{abstract}

%\pacs{
%xxx
%}

\maketitle

In chemistry and biology, water is widely referred to as the universal solvent \cite{franks_water_2000, eisenberg_structure_2005}. As salts dissolve in water, the anomalously large dielectric permittivity of water promotes the solubilization of salt by screening interionic Coulomb interactions. At the same time, the dielectric response of water is influenced by the presence of dissolved salts \cite{bluh_dielektrizitatskonstanten_1924, hasted_dielectric_1948, haggis1952dielectric, harris_dielectric_1957, hasted_dielectric_1958, christensen_dielectric_1966, buchner1999dielectric}. Almost 100 years ago, it was found that the static dielectric permittivity of sodium chloride (NaCl) solution decreases as more salt is dissolved \cite{bluh_dielektrizitatskonstanten_1924}. Later, more sophisticated experiments revealed a nonlinear behavior in which dielectric decrement slows down at high solute concentrations \cite{hasted_dielectric_1948,christensen_dielectric_1966, buchner1999dielectric}. A theoretical explanation of this phenomenon was conceived soon after the first experiment. As stated in their dielectric saturation theory, Debye \cite{debye1929polar} and Sack \cite{sack1926dielektrizitatskonstante} envisioned the formation of hydration shells due to the tendency of water dipoles to be aligned along electric fields of dissociated ions. Debye further estimated that ionic electric fields are strong enough to saturate the polarizability of water molecules near the ions and therefore lower the dielectric response \cite{debye1929polar}. Because of its built-in physical intuition, dielectric saturation has been, to date, the most adopted theory to explain dielectric decrement in salt water \cite{sack1926dielektrizitatskonstante, stiles1982dielectric, rinne_dissecting_2014, seal_dielectric_2019, gorji2020static}.

The past half-century has witnessed significant progress in understanding water through principles of quantum mechanics and statistical physics \cite{laasonen__1993, 17PNAS-Chen, zhang_modeling_2021, o2022crumbling, schran2021machine}. This progress calls into question the dielectric saturation explanation. Indeed, consensus has been reached that the high dielectric permittivity of water is closely associated with correlated dipole fluctuations of water molecules on the underlying hydrogen(H)-bond network \cite{kirkwood_dielectric_1939, noauthor_molecular_1951, sharma_dipolar_2007}. However, this collective dipolar response is missing in the picture of dielectric saturation which mainly concerns the suppressed dielectric response of individual water molecules \cite{sack1926dielektrizitatskonstante, debye1929polar, rinne_dissecting_2014}. More disturbingly, based on classical electrodynamics, dielectric saturation is estimated to occur on water molecules that are a few angstroms away from ions \cite{debye1929polar}. The above length scale is comparable to the estimated de Broglie wavelength of electrons at room temperature \cite{kandelousi_local_2018}. Physical interactions at such length scales are governed by quantum mechanics rather than a classical description. In this regard, density functional theory (DFT)-based \cite{hohenberg64, kohn65} ab initio molecular dynamics (AIMD) \cite{85L-CPMD} provides an ideal framework to predict properties of liquids from quantum mechanical principles. Indeed, recent AIMD simulations found that polarizabilities of water molecules in ionic first hydration shells are only slightly different from that in neat water \cite{gaiduk_local_2017, rozsa_molecular_2020}, which contradicts the dielectric saturation hypothesis.

Due to the long-range nature of the dipole-dipole interaction and the disordered liquid structure, the prediction of dielectric response in water demands both a spatially extensive model containing many hundreds of water molecules and a simulation time beyond nanoseconds \cite{krishnamoorthy_dielectric_2021, zhang2016computing}. However, AIMD simulations of such large timescale and system size are simply not feasible using current computer architectures. Thus, to date, dielectric decrement has been mostly studied using molecular dynamics with classical force fields, and the effect of electronic polarizability has been neglected  \cite{rinne_dissecting_2014, seal_dielectric_2019, zasetsky_dielectric_2001, fuentes-azcatl_sodium_2016}.

Herein, we overcome the challenge by studying dielectric decrement by combining AIMD and deep neural networks (DNNs)  \cite{zhang_deep_2022, 18L-Zhang, zhang2018end}. The liquid structures of NaCl solutions are simulated by a DNN that explicitly incorporates long-range electrostatic interactions  \cite{zhang_deep_2022} with periodic simulation cells containing about 4000 water molecules.  Importantly, the potential is trained on DFT calculations based on the strongly constrained appropriately normed (SCAN) functional  \cite{15L-Sun, 16NC-Sun}. In addition, a second DNN \cite{zhang_deep_2020} is trained separately for centers of electronic orbitals, in terms of maximally localized Wannier functions \cite{marzari_maximally_1997}. Notably, this second DNN allows us to rigorously partition the electronic charge density into contributions from dipole moments of individual water molecules. The dual DNNs enable efficient computations of dielectric permittivity at the DFT accuracy. (See Supplemental Material \cite{support} for more details on this methodology.)

\begin{figure}
	\includegraphics[width=0.46\textwidth]{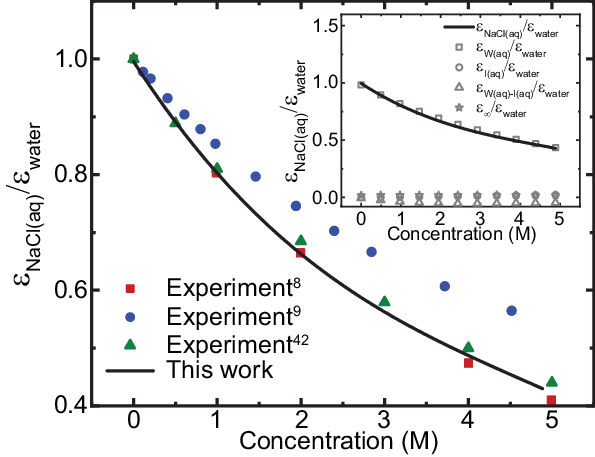}
	\caption{Static dielectric permittivity of NaCl solutions, $\varepsilon_{\rm{NaCl(aq)}}$, from this work and experiments \cite{christensen_dielectric_1966, buchner1999dielectric} \cz{\cite{shcherbakov2014dielectric}}. All results are normalized by the dielectric permittivity of neat water $\varepsilon_{\rm{water}}$. The inset shows the decomposition of the computed dielectric permittivity.}\label{fig1}
\end{figure}

Based on linear response theory, the static dielectric permittivity of NaCl solutions, $\text{\large{$\varepsilon$}}_\mathrm{NaCl(aq)}$, is related to the fluctuation of the total dipole moment, $\bm{\mathrm{M}}$, by \cite{sala_effects_2010, rinne_dissecting_2014}
\begin{flalign}  \label{eq1}
\text{\large{$\varepsilon$}}_\mathrm{NaCl(aq)} &=\frac{{\langle}\bm{\mathrm{M}}^{2}{\rangle}}{3 V k_{B} T \text{\large{$\varepsilon$}}_{0}}+\text{\large{$\varepsilon$}}_{\infty}&\notag\\
&=\frac{{\langle}(\bm{\mathrm{M}}_{\rm{W(aq)}}+\bm{\mathrm{M}}_{\rm{I(aq)}})^{2}{\rangle}}{3 V k_{B} T \text{\large{$\varepsilon$}}_{0}}+\text{\large{$\varepsilon$}}_{\infty}&\notag\\
        &=\text{\large{$\varepsilon$}}_\mathrm{W(aq)}+\text{\large{$\varepsilon$}}_\mathrm{W(aq)-I(aq)}+\text{\large{$\varepsilon$}}_\mathrm{I(aq)}+\text{\large{$\varepsilon$}}_{\infty}&&
\end{flalign}
where $V$, $k_{B}$, $T$, and $\text{\large{$\varepsilon$}}_{0}$ are the system volume, Boltzmann constant, temperature, and vacuum permittivity, respectively. $\text{\large{$\varepsilon$}}_{\infty}$ is the electronic contribution in the high-frequency limit. As expected, the theoretical $\text{\large{$\varepsilon$}}_{\infty}$ are small values around 1.88-1.99 at concentrations under consideration. We report the computed dielectric permittivity of NaCl solutions in Fig.~\ref{fig1} together with experimental data. Note that both results have been normalized to enable a better comparison of dielectric decrement behavior. There is good agreement between experiments and present calculations. In particular, the nonlinear behavior in dielectric decrement observed in experiment is well reproduced. The dielectric permittivity drops steeply at low concentrations, but its slope becomes gradually flattened as solute concentration increases. Notably, the nonlinearity generates a bowing feature in dielectric decrement. Absolute values of the computed dielectric permittivity are reported in Supplemental Material Table 1 \cite{support}.  It should be noted that the predicted dielectric permittivity of neat water by SCAN functional is 102.5, which is larger than the experimental value of 78. \cz{The overestimation of the dielectric permittivity is consistent with a previous study employing the SCAN functional \cite{krishnamoorthy_dielectric_2021}, and this overestimation} is particularly attributed to the self-interaction error in the SCAN functional that over-strengthens H-bonds. The slightly overstructured liquid water has been widely reported in literature \cite{17PNAS-Chen, xu_isotope_2020, piaggi2021phase} and its effects on observables can be approximated by the effects of decreasing the temperature, which does not affect our conclusions. 

In NaCl solutions, the fluctuation of the overall dipole moment, $\bm{\mathrm{M}}$, involves contributions from both water molecules, $\bm{\mathrm{M}}_{\rm{W(aq)}}$, and ions, $\bm{\mathrm{M}}_{\rm{I(aq)}}$. Therefore, the dielectric permittivity, $\text{\large{$\varepsilon$}}_\mathrm{NaCl(aq)}$ in Eq.~\ref{eq1} is composed of the self-terms, $\text{\large{$\varepsilon$}}_\mathrm{W(aq)}$ and $\text{\large{$\varepsilon$}}_\mathrm{I(aq)}$ whose dipole fluctuations are restricted to water molecules and solvated ions only, and the cross-coupling term $\text{\large{$\varepsilon$}}_\mathrm{W(aq)-I(aq)}$ reflecting dipole fluctuations in water induced by the movements of ions or vice versa. The computed values of above terms are presented in the inset of Fig.~\ref{fig1}. Notably, $\text{\large{$\varepsilon$}}_\mathrm{NaCl(aq)}$ is dominated by $\text{\large{$\varepsilon$}}_\mathrm{W(aq)}$ at all concentrations, which agrees with previous findings  \cite{rinne_dissecting_2014, chandra_static_2000-1}. Thus, dielectric decrement observed in NaCl solutions is due to the weakened dielectric response of solvent water molecules.

The dielectric component $\text{\large{$\varepsilon$}}_\mathrm{W(aq)}$ due to solvent water can be further evaluated via the dipolar correlation formalism proposed by Kirkwood \cite{kirkwood_dielectric_1939} as
\begin{equation}\label{epsilon_gk}
\text{\large{$\varepsilon$}}_\mathrm{W(aq)}=\frac{\rho\mu^2 G_K}{3 k_{B} T \text{\large{$\varepsilon$}}_{0}},   
\end{equation}
where $\rho$ and $\mu$ denote water number density and average dipole moment per water molecule respectively, and $G_K$ is the so-called correlation factor that measures the total angular correlations among water dipoles. In polar liquids, $G_K$ is obtained by the integration of the dipolar correlation function as $G_K=\int\mathcal{C}(\bm{r})d\bm{r}=\frac{1}{N}\sum_{i=1}^N\sum_{j=1}^N \bm{\hat{\mu}}_i\cdot\bm{\hat{\mu}}_j$, where $\bm{\hat{\mu}}_i$ is the unit vector of the $i$th molecular dipole and $N$ is the number of water molecules. The dipolar correlation is defined as $\mathcal{C}(\bm{r})={\langle}\bm{d}(\bm{0})\cdot \bm{d}(\bm{r}){\rangle}$, accounting for the spatial correlation between the dipolar density as a function of distance, $\bm{r}$. Because of the discretized nature of water molecules, the dipolar density is defined as $\bm{d}(\bm{r})=\sum_{i=1}^N \bm{\hat{\mu}}_i \delta(\bm{r}-\bm{r}_i)$ with $\bm{r}_i$ denoting the position vector of the $i$th water molecule.

\begin{figure}
	\includegraphics[width=0.5\textwidth]{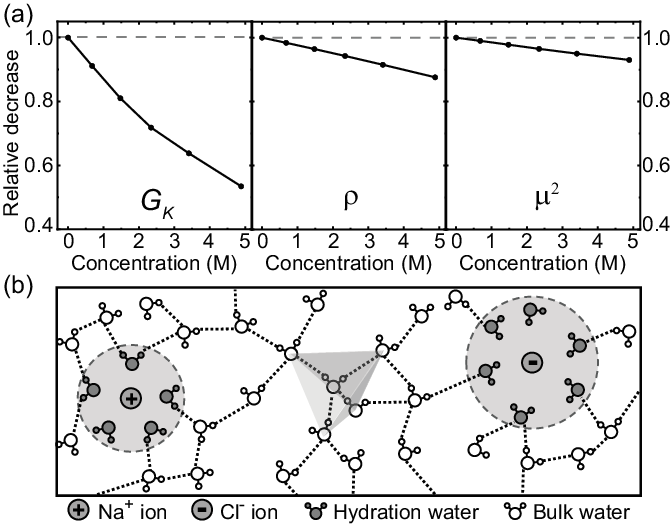}
	\caption{(a) Decrease of the correlation factor, $G_K$, water number density, $\rho$, and the square of water molecular dipole, $\mu{^2}$, as a function of concentration. All three quantities are normalized by their corresponding values at zero concentration. (b) Schematic diagram of the intrusion of ionic hydration shells into the tetrahedral H-bond network.}\label{fig2}
\end{figure}

In neat water, both the dipole moment, $\mu$, and the correlation factor, $G_K$, are largely enhanced by the underlying H-bond network, leading to the anomalously large dielectric permittivity \cite{sharma_dipolar_2007}. In NaCl solutions, as shown in Fig.~\ref{fig2} (a), the correlation factor, $G_K$, the water number density, $\rho$, and the water dipole moment, $\mu$, all decrease as increasing amounts of salt dissolved, which according to Eq.~\ref{epsilon_gk} leads to dielectric decrement.

\textit{\textbf{The effect from the disrupted H-bond network}}

\begin{figure*}
	\includegraphics[width=0.9\textwidth]{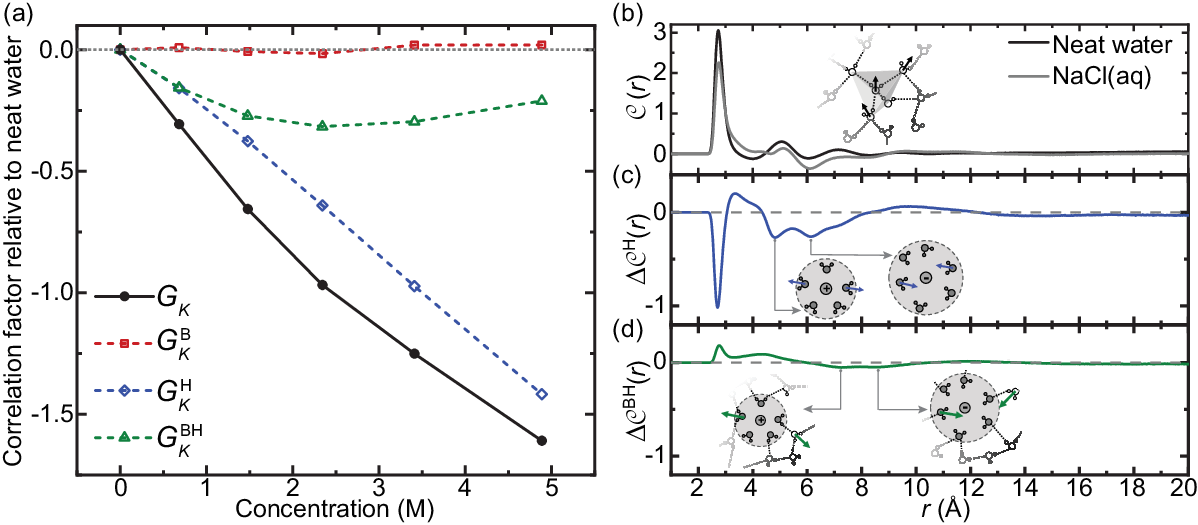}
	\caption{(a) Decomposition of the loss of correlation factor, $G_K$, of NaCl solutions relative to neat water into contributions from correlations between ``bulk water”, $G_K^{\rm{B}}$, ``hydration water”, $G_K^{\rm{H}}$, and cross-correlations between ``bulk water” and ``hydration water”, $G_K^{\rm{BH}}$. (b) The dipolar correlation function $\mathcal{C}(r)$ of neat water and 4.9 M NaCl solution. Decomposition of the difference of correlation factor, $\Delta \mathcal{C}(r)$, between 4.9 M NaCl solution and neat water into the contribution from (c) the correlation between ``hydration water” $\Delta \mathcal{C}^{\rm{H}}(r)$ and (d) the cross-correlation between ``bulk water” and ``hydration water”  $\Delta \mathcal{C}^{\rm{BH}}(r)$. Insets in (b-d) schematically show molecular configurations with arrows representing some representative molecular dipoles: H-bonded water dipoles in (b) point in similar directions contributing to a positive dipolar correlation; the presence of ions in (c) and (d) introduces negative dipolar correlations, as shown by the opposite directions of water molecular dipoles.}\label{fig3}
\end{figure*}

As seen in Fig.~\ref{fig2} (a), dielectric decrement of NaCl solutions is mostly attributed to the decreased correlation factor, $G_K$, relative to that of neat water. Thus, the strong correlation among dipole moments in neat water is significantly suppressed in salt solutions. In neat water, the large $G_K$ is closely associated with the tetrahedral H-bond structure, in which a water molecule at the center of a tetrahedron is H-bonded with four neighboring water molecules. The directions of dipole moments of any two H-bonded water molecules, therefore, point in a similar direction, resulting in a positive $\bm{\hat{\mu}}_i\cdot\bm{\hat{\mu}}_j$, which gives rise to the first positive sharp peak at 2.7 \r{A} in the dipolar correlation function in Fig.~\ref{fig3} (b). Under the influence of the directional H-bonding, dipole moments on vertices of a tetrahedron also prefer to be aligned in a similar direction to some extent, which yields a second positive peak around 5.1 \r{A} in Fig.~\ref{fig3} (b). In the same fashion, the dipolar correlation propagates to the third coordination shell and beyond.

The H-bond network is disrupted increasingly as more salt is dissolved. Salt ions exert electrostatic fields that can attract water molecules by competing with the H-bonding. In the close vicinity of ions, water molecules hydrate the ions by orienting their electric dipole moments towards the ions, thereby lowering the electrostatic energy of the system, as schematically shown in Fig.~\ref{fig2} (b). For a sodium cation, the first hydration shell can be described as a relatively tight sphere comprised of about 5 or 6 water molecules, whose oxygen is attractive to the cation at the center \cite{zhang_dissolving_2022}. On the other hand, the first hydration shell of a chloride ion is a relatively large sphere composed of as many as 6-8 water molecules whose protons are attracted to the chloride lone pair electrons  \cite{zhang_dissolving_2022, xu_importance_2021}. 

Because of the intrusion of the hydration shells, water molecules in the solvent are now divided into two distinct categories: the ``hydration (H) water” inside the ionic hydration shells and the ``bulk (B) water” outside it. As such, the pattern of dipolar correlation is fundamentally revised. As shown in Eq.~\ref{gk},
\begin{align} \label{gk}
G_K&=\int[\mathcal{C}^{\rm{B}}(\bm{r})+\mathcal{C}^{\rm{H}}(\bm{r})+\mathcal{C}^{\rm{BH}}(\bm{r})]d\bm{r}&\notag\\
        &=G_K^{\rm{B}}+G_K^{\rm{H}}+G_K^{\rm{BH}},&&
\end{align}
the total correlation factor $G_K$ involves the self-terms of $G_K^{\rm{B}}$ ($G_K^{\rm{H}}$) by dipolar correlation restricted to ``bulk water” (``hydration water”) only, and the coupling term $G_K^{\rm{BH}}$ due to the dipolar correlation between ``bulk water” and ``hydration water”. The above components in correlation factors, relative to neat water, are presented in Fig.~\ref{fig3} (a). (See: Supplemental Material \cite{support} for more details.)

As seen in Fig.~\ref{fig3} (a), the reduction in the overall correlation factor, $G_K$, is mostly from $G_K^{\mathrm{H}}$, which describes the correlation among ``hydration water”. This is because water molecules in hydration shells are constricted by the ion-water attraction instead of H-bonding. Within a hydration shell, the cation (anion)-water attraction reorientates the dipole moments from an H-bonding direction to a central-force direction pointing outwards (towards) ions. As such, the dipolar correlation between two neighboring ``hydration water” molecules is thereby significantly suppressed. This is evidenced by the sharp negative peak at $\sim$ 2.7 \r{A} in the dipolar correlation function $\Delta \mathcal{C}^{\rm{H}}(r)$ as plotted relative to neat water in Fig.~\ref{fig3} (c).  Moreover, the absence of H-bonding even causes anti-correlations between two ``hydration water” molecules located on the opposite sides of a single ion as schematically shown by opposite directions of water molecular dipoles in the inset of Fig.~\ref{fig3} (c). Therefore, the aforementioned positive peak of neat water in Fig.~\ref{fig3} (b) due to correlated dipole moments on vertices of a tetrahedron at 5.1 \r{A} disappears. Instead, it is replaced by two negative peaks at 4.8 and 6.1 \r{A}, which are caused by the anti-correlated water dipoles in hydration shells of Na$^+$ and Cl$^-$ ions, respectively. At long range, water molecules in a hydration shell, in principle, should be correlated to those in another hydration shell. However, such correlations are also weaker than those in neat water as expected in Fig.~\ref{fig3} (c). As concentration increases, the loss of $G_K^{\mathrm{H}}$ should accumulate linearly, which is responsible for most of the \textit{linear dielectric decrement} in salt water.

Of course, ``hydration water” is H-bonded to ``bulk water”, and in this way, the H-bond network is partially restored. Nevertheless, the reconstructed H-bond structure deviates from that found in neat water. Within a hydration shell, two water molecules located on opposite sides of a single ion are anti-correlated, as mentioned above. Because of the highly directional nature of H-bonding, the anti-correlation extends to the correlation between one ``hydration water” molecule and one ``bulk water” molecule that is H-bonded to another ``hydration water” molecule at the other side of the ion, as schematically shown by the opposite direction of green arrows in the inset of Fig.~\ref{fig3} (d). Again, these anticorrelations can be identified as a broad negative peak centered at 8 \r{A}, which weakens the dipolar correlation. As a result, $G_K^{\mathrm{BH}}$ also contributes to the decreased overall correlation factor of $G_K$ relative to neat water, as shown in Fig.~\ref{fig3} (a). Moreover, $G_K^{\mathrm{BH}}$ plays a surprisingly key role in the \textit{nonlinear dielectric decrement} as evidenced by its arc shape in Fig.~\ref{fig3} (a). This nonlinearity is an intrinsic property because $G_K^{\mathrm{BH}}$ describes the correlation between the dipolar density of ``bulk water” $\bm{d}^{\mathrm{B}}(\bm{r})$ and the dipolar density of ``hydration water” $\bm{d}^{\mathrm{H}}(\bm{r})$, and its value depends on the existence of both types of water, \textit{i.e.,} ${\langle}\bm{d}^{\mathrm{B}}(\bm{0})\cdot \bm{d}^{\mathrm{H}}(\bm{r}){\rangle}$. In neat water, $G_K^{\mathrm{BH}}=0$ since the dipolar density of ``hydration water” $\bm{d}^{\mathrm{H}}(\bm{r})$ is zero. As salt dissolve in water, hydration shells appear in the solution, and the absolute value of $G_K^{\mathrm{BH}}$ starts to increase, reaching its maximum at about 2.3 M, in which the NaCl solution is roughly equally occupied by ``bulk water” and ``hydration water”. After the maximum, $G_K^{\mathrm{BH}}$ decreases with further elevated concentrations. In principle, it will vanish again at $\bm{d}^{\mathrm{B}}(\bm{r})=0$, when the entire solution is completely occupied by hydration shells. 

The tetrahedral H-bond network is expected to recover in the ``bulk water” outside the hydration shell. The dipolar correlation among ``bulk water” molecules is captured by the $G_K^{\mathrm{B}}$ component of the correlation factor. Indeed, the analysis in Fig.~\ref{fig3} (a) shows that $G_K^{\mathrm{B}}$ of NaCl solutions at all concentrations is little different from neat water. Thus, the large decrease in the correlation factor, $G_K$, in salt water is mostly due to the disrupted H-bond network in the ``hydration water”.

\textit{\textbf{Excluded volume effect} }
Due to short-range repulsion, ions and water molecules are separated by 2-4 \r{A}. This extra volume demanded by ions is no longer accessible to water molecules, and the water number density is therefore decreased. In the literature, this is referred to as the excluded volume effect  \cite{adar_dielectric_2018}. According to Eq.~\ref{epsilon_gk}, this effect should lead to the decreased dielectric permittivity. Indeed, the present computations show that the excluded volume effect makes a small contribution to dielectric decrement, in which the water number density decreases slightly with increasing solute concentration as shown in Fig.~\ref{fig2} (a). Since the repelled volume by ions is proportional to the salt concentration, dielectric decrement due to the excluded volume effect is indeed linear as expected.

\textit{\textbf{Local field effect}}
Hydrated ions, like all charged defects, change the electrostatic potential profile throughout the solution. As expected, water molecules nearby an ion are polarized in a different manner from neat water. In condensed matter physics, related phenomena have been already identified, for example around defects in semiconductors or at interfaces in solid materials, and they have long been recognized as the local field effect \cite{kandelousi_local_2018}. There is consensus that a proper description of local field effects, particularly for regions close to charged defects, demands electronic structure details computed from quantum mechanics. Based on DFT, the present DNN simulations yield a dipole moment, $\mu$=2.85 (2.91) Debye for the ``hydration water” of the cation (anion), which is only slightly smaller than the value of 2.99 Debye in neat water. This suggests that the capability of ions to polarize the water dipole is comparable to that of H-bonding. Indeed, it is also consistent with the recent theoretical discovery that molecular polarizabilities of the ``hydration water” are only marginally different from that in neat water \cite{gaiduk_local_2017, rozsa_molecular_2020}. Since H-bonding is mostly electrostatic in nature, it strongly indicates that water molecules nearby ions are far from being saturated by ions’ local fields. Nevertheless, the local field effect also contributes slightly to dielectric decrement as indicated by Eq.~\ref{epsilon_gk}. Because the $\mu$ of the ``hydration water” is only a little smaller than in neat water, $\mu^2$ of NaCl solutions drops slowly as a function of concentration, as shown in Fig.~\ref{fig2} (a).

\cz{In addition to the SCAN \textit{ab initio} simulations, we also simulated the dielectric permittivity using the classical OPC water model \cite{izadi_building_2014}. As shown in Supplemental Material \cite{support}, the results obtained using the OPC model agree well with those from the SCAN-DFT approach. A notable distinction between the OPC model and the SCAN-DFT model is that the OPC model is a rigid model with a fixed dipole moment of 2.48 D, indicating that the DFT approach is necessary for accurately capturing the local field effect.}

In conclusion, dielectric decrement, as a century-old problem, has been extensively studied over decades. 
\cz{However, a critical question remains unresolved in the field regarding the main origin behind the dielectric decrement{\textemdash}whether it is the dielectric saturation effect \cite{debye1929polar, sack1926dielektrizitatskonstante} or the loss of dipolar correlation on the H-bond network \cite{rinne_dissecting_2014, seal_dielectric_2019}.
To provide an unambiguous answer, theoretical simulations must explicitly include both a polarizable model of water molecules and an accurate model of H-bonding, which can account for the dielectric saturation effect and correlation effect simultaneously. Importantly, the polarizable models of water molecules should be described from first principles at the quantum mechanics level, because the length scale of dielectric saturation effect is about a few angstroms which is comparable to the de Broglie wavelength of electrons at room temperature. In this work, we achieve the above goal by reproducing dielectric decrement in NaCl solutions on the DFT level using advanced DNNs. The results unambiguously determine that the dielectric decrement in NaCl solutions is dominated by the loss of correlations between water molecules due to the intrusion of ionic hydration shells into the H-bond network, while the contribution from dielectric saturation effect is small. 
Importantly, the present computations provide a quantitative explanation of dielectric decrement in salt water; we found that the linear dielectric decrement is due to the loss of correlation within hydration shells, while nonlinear dielectric decrement is due to the loss of correlation between water in hydration shells and bulk water.}

\begin{acknowledgments}
We thank Roberto Car, Linfeng Zhang, and Han Wang for fruitful discussions. This work was supported by National Science Foundation through Awards No. DMR-2053195. We also acknowledge support from the “Chemistry in Solution and at Interfaces” (CSI) Center funded by the U.S. Department of Energy through Award No. DE-SC0019394. This research used resources of the National Energy Research Scientific Computing Center (NERSC), which is supported by the U.S. Department of Energy (DOE), Office of Science under Contract No. DE-AC02-05CH11231. This research includes calculations carried out on HPC resources supported in part by the National Science Foundation through major research instrumentation grant number 1625061 and by the U.S. Army Research Laboratory under contract No. W911NF-16-2-0189. This research used resources of the Oak Ridge Leadership Computing Facility at the Oak Ridge National Laboratory, which is supported by the Office of Science of the U.S. Department of Energy under Contract No. DE-AC05-00OR22725. 
\end{acknowledgments}

\bibliography{reference}
\end{document}